\documentclass[twoside,fleqn]{article}
\usepackage{latexsym}
\usepackage{graphicx}
\usepackage{espcrc2}
\usepackage[usenames]{color}
\newcommand{\smallpicture}[1]{{\includegraphics[width=7cm]
{#1}}}

\newcommand{\tinypicture}[1]{{\includegraphics[width=3.6cm]
{#1}}}
\newcommand{\tinypictur}[1]{{\includegraphics[width=3.75cm]
{#1}}}
\definecolor{grey1}{rgb}{0.,0.,0.}
\definecolor{grey2}{rgb}{0.,0.,0.}
\definecolor{grey3}{rgb}{0.,0.,0.}
\newcommand{\tkl}{{{\color{grey3}
    T}{$\color{grey2}\chi$}{\color{grey1} L}}}

\newcommand{\bi}{\begin{itemize}}
\newcommand{\ei}{\end{itemize}}
\newcommand{\mpr}{\frac{m_{\pi}}{m_{\rho}}}
\newcommand{\beq}{\begin{equation}}
\newcommand{\eeq}{\end{equation}}
\newcommand{\dds}{\delta(\Delta S)}
\newcommand{\fig}[1]{Fig.~\ref{#1}}
\newcommand{\tab}[1]{Tab.~\ref{#1}}
\newcommand{\ks}{\kappa_{\rm sea}}
\newcommand{\kc}{\kappa_{\rm c}}

\newcommand{\kcv}{\kappa_{\rm c,v}}
\newcommand{\xpa}{\frac{\xi_{\pi}}{a}}

\newcommand{\IJMP}{Int.\ J.\ Mod.\ Phys.\ }

\title{Full QCD with dynamical Wilson
  fermions on a $24^{3}\times 40$-lattice -- \\a feasibility study\thanks{Talk presented by
    Th.~Lippert}}
\author{\tkl-Collaboration:\\ 
        L.~Conti$^{\rm a}$, N. Eicker$^{\rm b}$, L. Giusti$^{\rm a}$, 
        U.~Gl\"assner$^{\rm b}$, S.~G\"usken$^{\rm b}$, 
        H.~Hoeber$^{\rm c}$, Th.~Lippert$^{\rm c}$,
        G.~Martinelli$^{\rm a}$, F.~Rapuano$^{\rm a}$ 
        G.~Ritzenh\"ofer$^{\rm c}$,
        K.~Schilling$^{\rm b,c}$, G.~Siegert$^{\rm c}$
        A. Spitz$^{\rm b}$, and J. Viehoff$^{\rm c}$. \\[8pt]
{\rm $^a$}INFN, University "La Sapienzia", Roma,
          Italy,\\[8pt]
{\rm $^b$}Physics Department, University of Wuppertal, D-42097
           Wuppertal, Germany,\\[8pt]
{\rm $^c$}HLRZ c/o Forschungszentrum J\"ulich, D-52425 J\"ulich,
          and DESY, D-22603 Hamburg, Germany.}       
\begin{document}
\begin{abstract}
  The investigation of light sea-quark effects in lattice QCD with
  dynamical Wilson fermions requires both larger physical volumes and
  finer lattice resolutions than achieved previously.  As high-end
  supercomputers like the 512-node APE Tower provide the compute power
  to perform a major step towards the chiral limit (\tkl), we have
  launched a feasibility study on a $24^3 \times 40$ lattice.  We
  approach the chiral limit--while refining the resolution--, using
  the standard Wilson fermion action.  Following previous work, our
  Hybrid Monte Carlo simulation runs at $\beta=5.6$ and two
  $\kappa$-values, 0.1575 and 0.158.  From our study, we are confident
  that, for the APE Tower, a realistic working point has been found
  corresponding to a volume of 2 fm$^{3}$, with chirality
  characterized by $\frac{1}{{m}_{\pi}a} \approx 5.6$.
\end{abstract}
\maketitle


\section{INTRODUCTION}
Simulations of full QCD with Wilson fermions at zero temperature so
far have been carried out on lattices of size $\le 16^{3}\times 32$,
physical volumes $<$ 1.5 (fm)$^{3}$ and ratios of $\mpr > 0.6$
\cite{SESAM1}.  The latter quantity is a monitor for the closeness to
the chiral point.  It has been demonstrated \cite{GERO} that a
statistically significant full QCD (reference) sample can be generated
in one year's runtime on a 256-node APE computer, at $\mpr=0.71$.
This however corresponds still to a rather heavy quark mass $m_q$: by
use of chiral perturbation theory we can mock a fictitious "physical"
pseudoscalar meson containing two strange quarks, with mass ratio of
the size quoted, $ \frac{m_{ps}}{m_{\phi}}\approx
\frac{\sqrt{2m^{2}_{K}}}{m_{\phi}}=0.69$.  Thus, in order to quantify
light sea quark effects in full QCD, one would rather prefer to work
on larger volumes that accommodate a large $\pi$-correlation length both
in physical and lattice units. This clearly asks for simulations on lattices
$>16^3$ and $\beta\ge 5.6$.

In this note, we describe the \tkl\ project, which is geared to push
QCD simulations with standard Wilson fermions further {\em towards}
the {\em chiral} {\em limit}, i.e. beyond $\mpr<0.6$ and at
appropriate volumes.

The 512-node APE Tower offers sufficient memory to handle a
$24^3\times40$ lattice.  With its CPU-power it can drive an optimized
HMC at sufficient speed ({\em i}) to increase the lattice size by more
than a factor of 4 compared to the previous standards described above,
({\em ii}) to go more chiral, i.e., cope with worse conditioned
fermion matrices.

In our exploratory study, we are guided by the experiences described
in \cite{SESAM1,SESAM2}, taking advantage of algorithmic achievements
such as improved inverters \cite{FROMMER1} and new parallel
preconditioning techniques \cite{FROMMER2}. We shall report on a
Hybrid Monte Carlo simulation on a $24^3 \times 40$ lattice at
$\beta=5.6$ and two $\kappa$-values, 0.1575 and 0.158.

\section{HOW CHIRAL?}

The $24^3\times 40$ lattice allows to increase $\xi_{\pi}$ by a factor
of 1.5 compared to Ref.~\cite{SESAM3}, which should suffice to target
for $\mpr$ in the range of $.5$ -- $.6$.  Concerning physical volumes
and scales we benefit from the increasing $\Delta\beta$-shift
\cite{SESAM2} as we go to smaller bare quark masses and choose
$\beta=5.6$.

\begin{figure}[htb]
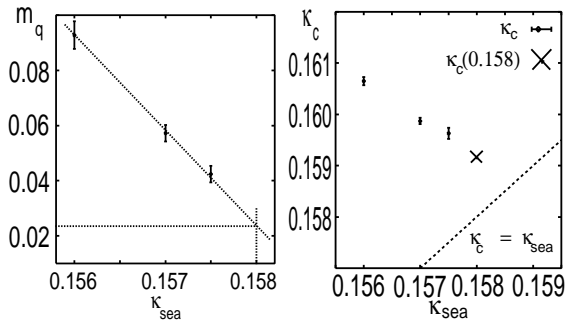

\tinypicture{find.eps}
\tinypictur{kappa_ex.eps}
\caption{
Fixing $\kappa_{\mbox{sea}}$ and
estimate for $\kcv$ compared to the results from Ref~\cite{SESAM3}.
\label{EXTRA}}
\end{figure}
For the determination of $\ks$, we extrapolated the relation
$m_qa=\frac{1}{2}\Big( \frac{1}{\kappa}- \frac{1}{\kc} \Big)$ on the
data set of Ref.~\cite{SESAM3} to $m_qa=0.023$, cf.\ \fig{EXTRA}a,
where $\frac{\xi_{\pi}}{24a}$ is estimated from the mass trajectory to
be about $.23$ of the spatial extension. This value is small enough to
protect us from finite size effects.  We  thus will work at
$\kappa=0.1580$.

To put this parameter choice into perspective, we sketched in
\fig{EXTRA}b the approach of the critical $\kappa$ at fixed $\ks$,
$\kcv$, \cite{SESAM3}, towards $\kc$, the locus of which is $\ks=\kc$
(the diagonal line). The cross marks the current estimate for our
working point. Notice that our parameter choice, $\ks=0.158$, appears
to be reasonably positioned within the `chirality gap'.

In the lattice discretization of the fermionic action, we switched
from the usual o/e representation to the full representation of
$M={\bf 1}-\kappa D$, employing a new SSOR preconditioning scheme
\cite{FROMMER2,FROMMER3,SESAM4}.  In particular on the APE machine,
this method offers an overall gain of 100 \% in execution time, as
seen from \tab{TIMES}.
\begin{table}[htb]
\footnotesize
\begin{tabular}{lllll}
\hline
Algo &      & $\kappa=0.1575$ & $\kappa=0.1580$   \\
\hline
o/e       &t/s   & 8200            & $-$               \\
SSOR      &t/s   & 3800            & 9100             \\
\hline
\vspace{8pt}
\end{tabular}
\caption{Average time to generate 1 trajectory  on the 
  APE100 Tower. \label{TIMES}}
\end{table}

\section{SIMULATION}

We have tuned the HMC timestep to achieve acceptance rates larger than
$60$ \%.  For the SSOR scheme with twice as many degrees of freedom as
in the o/e case, we chose $T=0.5$. For this trajectory length, in the
production runs, the 32 bit machine precision induces a reversibility
error $\dds$ in the range of just 2 \% of the average $\Delta S$ for
an inversion residue of $r= 10^{-8}$. This is due to {\em local}
computations, {\em global} summations being carried out in emulated
double precision arithmetic.  It should be said that the impact of
this error of $\Delta S$ onto the canonical distribution deserves
further attention.  The chosen HMC run parameters are given in
\tab{PARAMETER}.
\begin{table}[htb]
\footnotesize
\begin{tabular}{llllllll}
\hline
Algo&$\kappa$&$T$&$dt$ &acc/{\footnotesize\%} &$r$ \\
\hline
o/e      &$0.1575$&1  &0.008&70   &$10^{-8}$\\
SSOR     &$0.1575$&0.5&0.004&72   &$10^{-8}$\\
\hline
SSOR     &$0.1580$&0.5&0.004&66   &$10^{-8}$\\
\hline
\vspace{8pt}
\end{tabular}
\caption{Parameters of the HMC simulation.
\label{PARAMETER}}
\end{table}

During the thermalization phase we carefully approached the lowest
quark mass in a near adiabatic fashion, to protect the system from
oscillating through the shielding transition. We forked the run into
two $\kappa$-branches after an initial thermalization of 480
trajectories.  The production status reported is given by 830
trajectories (out of which 350 are thermalized) for $\kappa=0.1575$
and 1500 trajectories (out of which 750 are thermalized) for
$\kappa=0.1580$.

\section{PRELIMINARY RESULTS}

\begin{figure}[htb]
\smallpicture{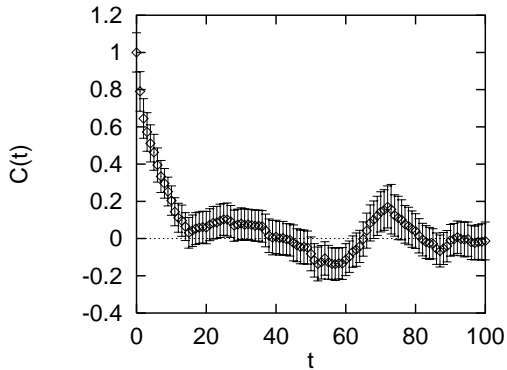}
\caption{
Autocorrelation function of plaquette.
\label{AUTO}}
\end{figure}

We have performed first tentative measurements of the autocorrelation
$C(t)$ of the plaquette.  In \fig{AUTO}, we plot the autocorrelation
function for the plaquette at $\kappa=0.158$.  On the large lattice
volume we can profit from a substantial self averaging effect
suppressing the fluctuations of the plaquette as well as of other
intensive quantities.  The plot indicates that the autocorrelation
times come out surprisingly small and might settle well below
$\tau_{\mbox{int}}=50$ for the plaquette.
We mention that the autocorrelation function for light meson masses
looks similar.

Using 22 configurations drawn from a sample of 600 trajectories we
have computed the potential following Ref.~\cite{SESAM2}.  We
performed a 30 step APE smearing and evaluated the potential for a
time extension of 5 where a plateau in the local mass is emerging.  We
have fitted for a "string tension" in the range up to 1 fm.  In
\tab{RESULTS} we quote a preliminary estimate for the ensueing scale
and physical lattice volume $V_{s}$.
\begin{table}[htb]
\footnotesize
\begin{tabular}{llllllll}
\hline
$a^{-1}$     & $V_{s}$  &$m_{\pi}$ SL &$m_{\rho}$ SL &$\mpr$  \\ 
\hline
$2.37(1)$ GeV   & 2 fm     & 0.178(5)    &0.32(2)       &0.56(4) \\
\hline
\vspace{8pt}
\end{tabular}
\caption{Results for the lattice scale from potential,
and masses in lattice units.
\label{RESULTS}}
\end{table}

We monitored local meson masses to position the run with respect to
chirality: on a sample of 19 configurations we retrieve a rough first
guesstimate of $m_{\pi}a$ and $m_{\rho}a$, see \tab{RESULTS}.  Our
findings suggest $\mpr$ to be 0.56(4).

\section{CONCLUSIONS AND OUTLOOK}

In the \tkl-feasibility study, we find that $\mpr$-ratio appears to
reach the target region indeed, where $\xpa=5.6<0.25 \times
V_s^{\frac{1}{3}}$. The lattice resolution is increased (with respect
to the small lattice results) to $a^{-1}=2.37$ GeV.  We are encouraged
by the observed autocorelation times and expect $>$ 50 independent
configurations from 8 months future runtime on APE Towers.
             
\section*{Acknowledgements}
We thank Prof. Mathis at ENEA/Italy and his staff for kind support.
We thank the Caspur group of La Sapienzia/Roma for help. Th. L.
and K. S. acknowledge the DFG-grant Schi 257/5-1.


\begin{thebibliography}{99}
\frenchspacing
%
\bibitem{SESAM1} SESAM-collab\-orat\-ion: SESAM-Collab\-orat\-ion,
  Gl\"assner et al, Nucl. Phys. {\bf B} {Proc. Suppl.} {\bf 47} (1996)
  386 and references quoted therein.
%
\bibitem{GERO} G. Ritzenh\"ofer, PhD-thesis, Wuppertal, 1996, to
  appear.
%
\bibitem{SESAM2} SESAM-Collaboration, Gl\"assner et al, Phys. Lett.
  {\bf B}, in print and talk presented by H. Hoeber, this volume.
%
\bibitem{FROMMER1} A.~Frommer, V.~Hannemann, B.~N\"ockel, Th.~Lippert,
  K.~Schilling, \IJMP {\bf C5} (1994) 1073.
%
\bibitem{FROMMER2} S. Fischer, A. Frommer, U. Gl\"a{}ssner, Th.
  Lippert, G. Ritzenh\"ofer and K. Schilling, Comp. Phys. Comm. {\bf
    1057} (1966) 1-15.
%
\bibitem{SESAM3} SESAM-collaboration, talk presented by U. Gl\"assner,
  this volume.
%
\bibitem{FROMMER3} Talk presented by A. Frommer, this volume.
%
\bibitem{SESAM4} SESAM-collaboration, talk presented by G.
  Ritzenh\"ofer, this volume.
%
\end{thebibliography}
\end{document}